%

\magnification\magstephalf
\def\ldt{\mathrel{.\,.}}
\def\[#1]{\hbox{\thickmuskip=4mu$[#1]$}}
\mathcode`\@="8000 {\catcode`\@=\active \gdef@{\mkern1mu}}
\baselineskip14pt
\parskip5pt
\def\proof{\smallskip\noindent{\bf Proof.}\quad}
\def\pfbox
  {\hbox{\hskip 3pt\lower2pt\vbox{\hrule
  \hbox to 9pt{\vrule height 7pt\hfill\vrule}
  \hrule}}\hskip3pt}
\def\bib[#1] {\smallskip\noindent\hangindent 20pt\hbox to20pt{[#1]\hfil}}
\def\AW{Addison\kern.1em--Wesley}

\def\DeM{1}
\def\DZ{2}
\def\Fii{3}
\def\FF{4}
\def\Gold{5}
\def\Kii{6}
\def\Kiii{7}
\def\KMP{8}
\def\Lind{9}
\def\Poin{10}
\def\Yao{11}

\centerline{\bf Shellsort With Three Increments}
\centerline{by Svante Janson and Donald E. Knuth}

\bigskip
{\narrower\noindent\smallskip\noindent
{\bf Abstract.} A perturbation technique can be used to simplify and
sharpen A.~C. Yao's theorems about the behavior of shellsort with increments
$(h,g,1)$. In particular, when $h=\Theta(n^{7/15})$ and
$g=\Theta(h^{1/5})$, the average running time is $O(n^{23/15})$. The proof
involves interesting properties of the inversions in random permutations
that have been $h$-sorted and $g$-sorted.
\smallskip}

\medskip
\noindent
Shellsort, also known as the ``diminishing increment sort''
[\Kiii, Algorithm 5.2.1D],
puts the elements of an array $(X_0,\ldots, X_{n-1})$ into order by
successively performing a straight insertion sort on larger and larger
subarrays of equally spaced elements. The algorithm consists of $t$~passes
defined by increments $(h_{t-1},\ldots,h_1,h_0)$, where $h_0=1$; the
$j$\/th pass makes $X_k\leq X_l$ whenever $l-k=h_{t-j}$.

A. C. Yao
[\Yao]
 has analyzed the average behavior of shellsort in the
general three-pass case when the increments are $(h,g,1)$. The most
interesting part of his analysis dealt with
 the third pass, where the running time
is $O(n)$ plus a term proportional to the average number of inversions that
remain after a random permutation  has been $h$-sorted and $g$-sorted. Yao
proved that if $g$ and~$h$ are relatively prime, the average number of
inversions remaining~is
$$\psi(h,g)@n+\widehat{O}(n^{2/3})\,,\eqno(0.1)$$
where the constant implied by $\widehat O$ depends on $g$ and~$h$. He gave a
complicated triple sum for $\psi(h,g)$, which is too difficult to explain
here; we will show that
$$\psi(h,g)={1\over 2}\,\sum_{d=1}^{g-1}\,\sum_r\,{h-1\choose
r}\left({d\over g}\right)^{\!r}\left(1-{d\over g}\right)^{\!h-1-r}\,
\left|\, r-\left\lfloor{hd\over g}\right\rfloor\right|\,.\eqno(0.2)$$
Moreover, we will prove that the average number of inversions after such
$h$-sorting and $g$-sorting~is
$$\psi(h,g)@n+O(g^3h^2)\,,\eqno(0.3)$$
where the constant implied by $O$ is independent of $g$, $h$, and~$n$.

The main technique used in proving (0.3) is to consider a stochastic
algorithm ${\cal A}$ whose output has the same distribution as the
inversions of the third pass of shellsort. Then by slightly perturbing the
probabilities that define~${\cal A}$, we will obtain an algorithm ${\cal
A}^{\ast}$ whose output has the expected value $\psi(h,g)@n$ exactly.
Finally we will prove that the perturbations cause the expected value to
change by at most $O(g^3h^2)$.

Section 1 introduces basic techniques for inversion counting, and section~2
adapts those techniques to a random input model. Section~3 proves that
the crucial random variables needed for inversion counting are nearly
uniform; then
section~4 shows that the leading term $\psi(h,g)@n$ in (0.3) would
be exact if those variables were perfectly uniform. Section~5 explains how
to perturb them so that they are indeed uniform, and section~6 shows how
this perturbation yields the error term $O(g^3h^2)$ of~(0.3).

The asymptotic value of $\psi(h,g)$ is shown to be $(\pi h/128)^{1/2}g$ in
section~7. The cost of the third pass in
$(ch,cg,1)$-shellsort for $c>1$ is analyzed in section~8. This makes it
possible to bound the total running time for all three passes, as shown in
section~9, leading to an $O(n^{23/15})$ average running time when $h$
and~$g$ are suitably chosen.

The bound $O(g^3h^2)$ in (0.3) may not be best possible. Section~10
discusses a conjectured improvement, consistent with computational
experiments, which would reduce the average cost to $O(n^{3/2})$ if it
could be proved.

The tantalizing prospect of extending the techniques of this paper to more
than three increments is explored briefly in section~11.

\bigskip\noindent
{\bf 1. Counting inversions.}\quad
We shall assume throughout this paper that $g$ and $h$ are relatively
prime. To fix the ideas, suppose $h=5$, $g=3$, $n=20$, and suppose we are
sorting the 2-digit numbers
$$(X_0,X_1,\ldots,X_{n-1})= (03, 14, 15, 92, 65,
35, 89, 79, 32, 38, 46, 26, 43, 37, 31, 78, 50, 28, 84, 19)\,.$$
(Cf.\
[\Kii, Eq.\ 3.3--(1)].)
The first pass of shellsort, $h$-sorting,
replaces this array by
$$(X'_0,X'_1,\ldots,X'_{n-1})=  (03, 14, 15, 32,
19, 35, 26, 28, 37, 31, 46, 50, 43, 84, 38, 78, 89, 79, 92, 65)\,.$$
The second pass,
$g$-sorting, replaces it by
$$(X''_0,X''_1,\ldots,X''_{n-1})=
(03, 14, 15, 26, 19, 35, 31, 28, 37, 32, 46, 38, 43, 65, 50, 78, 84,
79, 92, 89)\,.$$
Our task is to study the inversions of this list, namely the pairs $k$,~$l$
for which $k<l$ and $X''_k>X''_l$.

The result of $g$-sorting is the creation of $g$ ordered lists
$X''_j<X''_{j+g}<X''_{j+2g}<\cdots$ for $0\leq j<g$, each of which contains
no inversions within itself. So the inversions remaining are inversions
between different sublists. For example, the 20~numbers sorted above
lead~to
$$\vcenter{\halign{#\hfil\cr
list $0=(03,26,31,32,43,78,92)\,,$\cr
list $1=(14,19,28,46,65,84,89)\,,$\cr
list $2=(15,35,37,38,50,79)\,;$\cr}}$$
the inversions between list 0 and list 1 are the inversions of
$$(03,14,26,19,31,28,32,46,43,65,78,84,92,89)\,.$$
It is well known
[\Kiii, \S5.21]
that two interleaved ordered lists of length $m$ have $\sum_{r=0}^{m-1}
|r-s_r|$ inversions, where $s_r$ of the elements of the second
list are less than the $(r+1)$\/st element of the first list; for example,
$(03,14,26,\ldots,89)$ has
$$|0-0|+|1-2|+|2-3|+|3-3|+|4-3|+|5-5|+|6-7|=4$$
inversions.
If $r\geq s_r$, the $(r+1)$\/st element of the first list is inverted by
$r-s_r$ elements of the second; otherwise it inverts $s_r-r$ of
those elements. (We assume that the list elements are distinct.)
The same formula holds for interleaved ordered lists of lengths $m$
and~$m-1$, because we can imagine an infinite element at the end of the second
list.

Let $Y_{kl}$ be the number of elements $X_{k'}$ such that
$k'\equiv k$ (mod~$h$) and
$X_{k'}<X_l$. The $n$ numbers $Y_{ll}$ for $0\leq l<n$
clearly characterize the permutation performed by $h$-sorting; and it is
not hard to see that the full set of $hn$ numbers~$Y_{kl}$ for $0\leq k<h$
and $0\leq l<n$ is enough to
determine the relative order of all the $X$'s.

There is a convenient way to enumerate the inversions that remain after
$g$-sorting, using the numbers $Y_{kl}$. Indeed, let
$$J_{kl}=(k\bmod h +hY_{kl})\,\bmod g\,.\eqno(1.1)$$
Then $X_l$ will appear in list $j=J_{ll}$ after $g$-sorting. Let $S_{jl}$
be the number of elements~$X_{k'}$ such that
$X_{k'}<X_l$ and $X_{k'}$ is in list~$j$. The inversions between lists~$j$
and~$j'$ depend on the difference $|S_{jl}-S_{j'l}|$ when $X_l$
goes into list~$j$.

Given any values of $j$ and $j'$ with $0\leq j<j'<g$, let
$j_s=(j+hs)\bmod g$, and let $d$ be minimum with $j_d=j'$. Thus, $d$~is
the distance from $j$ to~$j'$ if we count by steps of $h\;{\rm modulo}\;
g$.  Let
$$H=\{j_1,j_2,\ldots,j_d\}\eqno(1.2)$$
be the
$h$~numbers between $j$ and $j'$
 in this counting process, and let $Q_l$ be the number of indices~$k$ such
that $0\leq k<h$ and $J_{kl}\in H$. Then we can prove the following basic
fact:

\proclaim
Lemma 1. Using the notation above, we have
$$S_{jl}-S_{j'l}=Q_l-\lfloor hd/g\rfloor\eqno(1.3)$$
for all $j$, $j'$, and $l$ with $0\leq j<j'<g$ and $0\leq l<n$.

\proof Since the $X$'s are distinct, there is a permutation
$(l_0,l_1,\ldots,l_{n-1})$ of $\{0,1,\ldots,n-1\}$ such that
$X_{l_0}<X_{l_1}<\cdots <X_{l_{n-1}}$. We will prove (1.3) for $l=l_t$ by
induction on~$t$.

Suppose first that $l=l_0$, so that $X_l$ is the smallest element being
sorted. Then $Y_{kl}=0$ for all~$k$; hence $J_{kl}=k\bmod g$ for $0\leq
k<h$. Also $S_{jl}=S_{j'l}=0$. Therefore (1.3) is equivalent in this case
to the assertion that {\sl precisely\/} $\lfloor hd/g\rfloor$ {\sl elements
of the multiset\/}
$$\{0\bmod g,\;1\bmod g,\;\ldots,\;(h-1)\bmod g\}$$
{\sl belong to\/} $H$.

A clever proof of that assertion surely exists, but what is it? We can at
any rate use brute force by assuming first that $j=0$. Then the number of
solutions to $x\equiv hd$ (mod~$g$) and $0\leq x<h$ is the number of
integers in the interval $\bigl[-hd/g \ldt -h(d-1)/g\bigr)$, namely
$\lceil -h(d-1)/g\rceil-\lceil -hd/g\rceil=\lfloor hd/g\rfloor-\lfloor
h(d-1)/g\rfloor$. Therefore the assertion for $j=0$ follows by induction
on~$d$. And once we've proved it for some pair $j<j'$, we can prove it for
$j+1<j'+1$, assuming that $j'+1<g$: The value of~$d$ stays the same, and
the values of $j_1,j_2,\ldots,j_d$ increase by~1 (mod~$g$).
 So we lose one
solution if $j_s\equiv h-1$ (mod~$g$) for some~$s$ with
$1\leq s\leq d$; we
gain one solution if $j_s\equiv -1$ (mod~$g$) for
some~$s$. Since $j_s\equiv
h-1\Longleftrightarrow j_{s-1}\equiv -1$, the net change is zero unless
$j_1\equiv h-1$ (but then $j=g-1$) or $j_d\equiv -1$ (but then $j'=g-1$).
This completes the proof by brute force when $l=l_0$.

Suppose (1.3) holds for $l=l_t$; we want to show that it also holds
when $l$ is replaced by $l'=l_{t+1}$. The numbers $Y_{kl}$ and $Y_{kl'}$
are identical for all but one value of~$k$, since
$$Y_{kl'}=Y_{kl}+\[l\equiv k\;({\rm mod}\;h)]\,.$$
Thus,
the values of $J_{kl}$ and $J_{kl'}$ are the same except that $J_{kl}$
increases by $h$ (mod~$g$) when $k\equiv l$ (mod~$h$). It
follows that
$$Q_{l'}=Q_l+\[J_{ll}=j]-\[J_{ll}=j']\,.$$
This completes the proof by induction on $t$, since
$S_{jl'}=S_{jl}+\[J_{ll}=j]$ for all~$j$.~~\pfbox

\proclaim
Corollary. Using the notations above, the total number of inversions
between lists~$j$ and~$j'$~is
$$\sum_{l=0}^{n-1}\,\bigl|\, Q_l-\lfloor hd/g\rfloor\,\bigr|
\,\[J_{ll}=j]\,.
\eqno(1.4)$$

\proof
This is $|S_{jl}-S_{j'l}|=|r-s_r|$ summed over all $r$ such
that $X_l$ is the $(r+1)$\/st element of list~$j$.~~\pfbox

\smallskip
In the example of $n=20$ two-digit numbers given earlier, with $h=5$,
$g=3$, $j=0$,  and $j'=1$, we have $d=2$, $H=\{2,1\}$,
$$\vcenter{\halign{$\hfil#\;$%
&$\hfil#$\quad&$\hfil#$\quad&$\hfil#$\quad&$\hfil#$\quad&$\hfil#$\quad
&$\hfil#$\quad&$\hfil#$\quad&$\hfil#$\quad&$\hfil#$\quad&$\hfil#$\quad
&$\hfil#$\quad&$\hfil#$\quad&$\hfil#$\quad&$\hfil#$\quad&$\hfil#$\quad
&$\hfil#$\quad&$\hfil#$\quad&$\hfil#$\quad&$\hfil#$\quad&$\hfil#$\cr
l=&0&1&2&3&4&5&6&7&8&9&10&11&12&13&14&15&16&17&18&19\cr
\noalign{\smallskip}
X_l=&03&14&15&92&65&35&89&79&32&38&46&26&43&37&31&78&50&28&84&19\cr
Y_{0l}=&0&1&1&4&3&1&4&4&1&2&2&1&2&2&1&3&3&1&4&1\cr
Y_{1l}=&0&0&1&4&3&2&3&3&2&2&2&1&2&2&2&3&2&2&3&1\cr
Y_{2l}=&0&0&0&4&3&2&4&3&2&2&3&1&2&2&2&3&3&1&4&1\cr
Y_{3l}=&0&0&0&3&2&1&3&2&0&2&2&0&2&1&0&2&2&0&2&0\cr
Y_{4l}=&0&0&0&4&3&2&4&4&2&2&3&1&3&2&1&4&3&1&4&0\cr
J_{0l}=&\underline{0}&2&2&2&0&\underline{2}&2&2&2&1&\underline{1}%
&2&1&1&2&\underline{0}&0&2&2&2\cr
J_{1l}=&1&\underline{1}&0&0&1&2&\underline{1}&1&2&2&2&\underline{0}%
&2&2&2&1&\underline{2}&2&1&0\cr
J_{2l}=&2&2&\underline{2}&1&2&0&1&\underline{2}&0&0&2&1&\underline{0}%
&0&0&2&2&\underline{1}&1&1\cr
J_{3l}=&0&0&0&\underline{0}&1&2&0&1&\underline{0}&1&1&0&1&\underline{2}%
&0&1&1&0&\underline{1}&0\cr
J_{4l}=&1&1&1&0&\underline{1}&2&0&0&2&\underline{2}&1&0&1&2&\underline{0}%
&0&1&0&0&\underline{1}\cr
Q_l=&3&4&3&2&4&4&3&4&3&4&5&2&4&4&2&3&4&3&4&3\cr
}}$$
and the underlined values $J_{ll}$ are 0 for $l=$ 0, 3, 8, 11, 12, 14, 15
(accounting for the seven elements in list~0). The inversions between
lists~0 and~1 are therefore
$$\postdisplaypenalty10000
|3-3|+|2-3|+|3-3|+|2-3|+|4-3|+|2-3|+|3-3|=4$$
according to (1.4).

\medbreak\noindent
{\bf 2. Random structures.}\quad
We obtain a random run of shellsort if we assume that the input array
$(X_0,X_1,\ldots,X_{n-1})$ is a random point in the $n$-dimensional unit
cube. For each integer~$l$ in the range $0\leq l<n$ and for each ``time''~$t$
in the range $0\leq t\leq 1$, we will consider the contribution made
by~$X_l$ to the total number of inversions if $X_l=t$.

Thus, instead of the quantities $Y_{kl}$ and $J_{kl}$ defined in the
previous section, we define
$$\eqalignno{%
Y_{kl}(t)&=\sum_{\scriptstyle k'\equiv k\,(\bmod\;h)
\atop\scriptstyle 0\leq k'<n}\;\[X_{k'}<t]\,,&(2.1)\cr
\noalign{\smallskip}
J_{kl}(t)&=\bigl(k\bmod h+hY_{kl}(t)\bigr)\,\bmod\;g\,.&(2.2)\cr}$$
These equations are almost, but not quite, independent of $l$, because we
assume that $X_l=t$ while all other $X$'s are uniformly and independently
random.

For each pair of indices $j$ and $j'$ with $0\leq j<j'<g$, we define $H$ as
in (1.2), and we let
$$Q_l(t)=\sum_{k=0}^{h-1}\,\[J_{kl}(t)\in H]\[k\neq l\bmod h]\,.\eqno(2.3)$$
This definition is slightly different from our original definition
of~$Q_l$, because we have excluded the term for $k=l\bmod h$. However,
formula~(1.4) remains valid because $j\notin H$; when $J_{ll}=j$, the
excluded term is therefore zero.

{\tolerance=5000
Notice that, for fixed $l$, the random variables $Y_{kl}(t)$ for $0\leq
k<h$ are independent. Therefore the random variables $J_{kl}(t)$ are
independent; and $Q_l(t)$ is independent of~$J_{ll}(t)$.
The average contribution of~$X_l$ to the inversions between lists~$j$
and~$j'$ when $X_l=t$ is therefore
$$W_{jj'l}(t)=\Pr\[J_{ll}(t)=j]\,{\rm E}\,\bigl|Q_l(t)-\lfloor
hd/g\rfloor\bigr| \eqno(2.4)$$
by (1.4), where probabilities and expectations are computed with respect to
$(X_0,\ldots,X_{l-1},\allowbreak X_{l+1},\ldots,X_{n-1})$. The average
total contribution of~$X_l$ is obtained by integrating over all
values of~$t$:
\par}

\proclaim
Lemma 2. Let
$$W_{jj'l}=\int_0^1W_{jj'l}(t)\,dt\,.\eqno(2.5)$$
Then the average grand total number of inversions in the third pass of
shellsort~is
$$\sum_{\scriptstyle 0\leq j<j'<g\atop\scriptstyle 0\leq l<n}\,
W_{jj'l}\,.\quad\pfbox\eqno(2.6)$$
{\rm
Our goal is to find the asymptotic value of this sum, by proving that it
agrees with the estimate (0.3) stated in the introduction.}

\medbreak\noindent
{\bf 3. Near uniformity.}\quad
The complicated formulas of the previous section become vastly simpler when
we notice that each random variable $J_{kl}(t)$ is almost uniformly
distributed: The probability that $J_{kl}(t)=j$ is very close to~$1/g$, for
each~$j$, as long as $t$ is not too close to~0 or~1. To prove this
statement, it suffices to show that $Y_{kl}(t)\bmod g$ is approximately
uniform, because $h$ is relatively prime to~$g$. Notice that $Y_{kl}(t)$
has a binomial distribution, because it is the sum of approximately $n/h$
independent random 0--1 variables that take the value~1 with
probability~$t$.

\proclaim
Lemma 3. If\/ $Y$ has the binomial distribution with parameters $(m,t)$, then
$$\left|\Pr\[Y\bmod g=j]-{1\over g}\right|
< {1\over g}\,\phi_{gm}(t)\eqno(3.1)$$
for $0\leq j<g$, where
$$\phi_{gm}(t)=2\;\sum_{k=1}^{\infty}\,e^{-8t(1-t)k^2m/g^2}\,.\eqno(3.2)$$

\proof
Let $y_j=\Pr\[Y\bmod g=j]$, and consider the discrete Fourier
transform
$$\hat{y}_k=\sum_{j=0}^{g-1}\,\omega^{kj}y_j={\rm E}\,\omega^{kY}$$
where $\omega =e^{2\pi i/g}$. We have
$$\hat{y}_k=\sum_{l=0}^m\,{m\choose l}t^{@l}(1-t)^{m-l}\omega^{kl}=
(\omega^k t+1-t)^m\,,\eqno(3.3)$$
and
$$\eqalignno{|@\omega^kt+1-t@|^2
&=t^2+(1-t)^2+t(1-t)(\omega^k+\omega^{-k})\cr
&=1-2t(1-t)(1-\cos 2\pi k/g)\cr
&=1-4t(1-t)\sin^2\pi k/g\,.&(3.4)\cr}$$
If $0\leq x\leq \pi/2$ we have $\sin x\geq 2x/\pi$; hence, if $0\leq k\leq
{1\over 2}g$,
$$|@\omega^kt+1-t@|^2\leq 1-
16t(1-t)k^2\!/g^2<e^{-16t(1-t)k^2\!/g^2}\,.$$
And if ${1\over 2}g<k<g$ we have $|\hat{y}_k|
=|\hat{y}_{g-k}|$. Therefore
$$\sum_{k=1}^{g-1}\,|\hat{y}_k|\leq
2\sum_{k=1}^{g/2}\,e^{-8t(1-t)k^2m/g^2} <\phi_{gm}(t)\,.\eqno(3.5)$$
The desired result follows since
$$y_j={1\over g}\,\sum_{k=0}^{g-1}\,\omega^{-kj}\hat{y}_k$$
and thus
$$\left|@y_j-{1\over g}\right|=\left|{1\over g}\,
\sum_{k=1}^{g-1}\,\omega^{-kj}\hat{y}_k@\right|\le
{1\over g}\,\sum_{k=1}^{g-1}\,|\hat{y}_k|\,.\quad\pfbox$$

\proclaim
Corollary. We have
$$\left|\Pr\[J_{kl}(t)=j]-{1\over g}\right|<{1\over
g}\,\phi(t)\eqno(3.6)$$
for $0\leq k<h$, where
$$\phi(t)=\cases{
2\sum_{k=1}^{\infty}e^{-4t(1-t)k^2n/g^2h}\,,&if $n\ge4h$;\cr
\noalign{\smallskip}
g,&if $n<4h$.\cr}\eqno(3.7)$$

\proof
Each variable $Y_{kl}(t)$ in (2.1) for $0\leq k<h$ has the binomial
distribution with parameters $(m,t)$, where if $n\ge4h$
$$m=\lceil(n-k)/h\rceil -\[k=l\bmod h]\geq {n\over h}-2\geq {n\over
2h}\,.$$
Now $J_{kl}(t)=j$ if and only if $Y_{kl}(t)$ has a certain value
mod~$g$. The case $n<4h$ is trivial.~~\pfbox

\medbreak\noindent
{\bf 4. Uniformity.}\quad
Let's assume now that, for given $l$ and $t$, the random variables
$J_{kl}(t)$ have a perfectly uniform distribution. Since the variables
$J_{kl}(t)$ are independent for $0\leq k<h$, this means that
$$\Pr\[J_{0l}(t)=j_0, J_{1l}(t)=j_1,\ldots,J_{(h-1)l}(t)=j_{h-1}]
={1\over g^h}\eqno(4.1)$$
for all $h$-tuples $(j_0,j_1,\ldots,j_{h-1})$.

In such a case the random variable $Q_l(t)$ defined in (2.3) is the
sum of $h-1$ independent indicator variables, each equal to 1
with probability $d/g$ because $H$ has $d$ elements.
Hence $Q_l(t)$ has the binomial distribution with parameters $(h-1,d/g)$,
and it is equal to~$r$ with probability
$${h-1\choose r}
\left({d\over g}\right)^{\!r}\left(1-{d\over g}\right)^{\!h-1-r}\,.\eqno(4.2)$$

Let $W^{\ast}_{jj'l}(t)$ be the value of $W_{jj'l}(t)$ under the assumption of
uniformity $\bigl($see (2.4)$\bigr)$. Thus $W^{\ast}_{jj'l}(t)$ is independent
of~$t$, and we let $W^{\ast}_{jj'l}=W^{\ast}_{jj'l}(t)$ in accordance
with~(2.5). Then
$$W^{\ast}_{jj'l}={1\over g}\,\sum_{r=0}^{h-1}\,{h-1\choose r}\left({d\over
g}\right)^{\!r} \left(1-{d\over g}\right)^{\!h-1-r}\left|\, r-\left\lfloor
{hd\over g}\right\rfloor\right|\,.\eqno(4.3)$$

For given values of $d$ and $j$, the index $j'=(j+hd)\,\bmod g$ is at
distance~$d$ from~$j$. Suppose that $a(d)$ of these pairs $(j,j')$ have $j<j'$.
Then $g-a(d)$ of them have $j>j'$, and $a(g-d)=g-a(d)$ since $j$ is at
distance $g-d$ from~$j'$. The sum of (4.3) over all $j<j'$ is therefore
independent of~$a(d)$:
$$\eqalign{\sum_{0\leq j<j'<g}\,W^{\ast}_{jj'l}
&=\sum_{d=1}^{g-1}\,{a(d)\over g}\,\sum_{r=0}^{h-1}\,{h-1\choose
r}\left({d\over g}\right)^{\!r}\left(1-{d\over g}\right)^{\!h-1-r}
\left|\,r-\left\lfloor{hd\over g}\right\rfloor\right|\cr
\noalign{\smallskip}
&=\sum_{d=1}^{g-1}\,{a(g-d)\over g}\,\sum_{r=0}^{h-1}\,{h-1\choose
h-1-r}\left({g-d\over g}\right)^{\!h-1-r} \left(1-{g-d\over g}\right)^{\!r}\cr
\noalign{\smallskip}
&\hskip12em \times
\left|\,h-1-r-\left\lfloor{h(g-d)\over g}\right\rfloor\right|\cr
\noalign{\smallskip}
&={1\over 2}\,\sum_{d=1}^{g-1}\,\sum_{r=0}^{h-1}\,
{h-1\choose r}\left({d\over g}\right)^{\!r}
\left(1-{d\over g}\right)^{\!h-1-r}
\left|\,r-\left\lfloor{hd\over g}\right\rfloor\right|\,.\cr}$$
(We have used the fact that $\lfloor h(g-d)/g\rfloor =h-1-\lfloor
hd/g\rfloor$ when $hd/g$ is not an integer.) But this is just the quantity
$\psi(h,g)$ in (0.2), for each value of~$l$. We have proved

\proclaim
Lemma 4. If we assume that the variables $J_{kl}(t)$ have exactly the
uniform distribution, the quantity\/ $(2.6)$ is exactly $\psi(h,g)@n$.~~\pfbox

\medbreak\noindent
{\bf 5. Perturbation.}\quad
To complete the proof of (0.3), we use a general technique applicable to
the analysis of many algorithms: If a given complicated algorithm~${\cal
A}$ almost always has the same performance characteristics as a simpler
algorithm~${\cal A}^{\ast}$, then the expected performance of~${\cal A}$ is
the same as the performance of~${\cal A}^{\ast}$ plus an error term based
on the cases where ${\cal A}$ and ${\cal A}^{\ast}$ differ.
(See, for example, the analysis in
[\KMP],
where this ``principle  of negligible perturbation'' is applied to a
nontrivial branching process.)

{\tolerance=5000
In the present situation we retain the $(n-1)$-dimensional probability space
$(X_0,\ldots,X_{l-1},\allowbreak t,\allowbreak
X_{l+1},\ldots,X_{n-1})$ on which the random
variables $J_{kl}(t)$ were defined in (2.2), and we define a new set of
random variables $J^{\ast}_{kl}(t)$ on the same space, where
$J^{\ast}_{kl}(t)$ has exactly a uniform distribution on
$\{0,1,\ldots,g-1\}$. This can be done in such a way that
$J_{kl}(t)=J^{\ast}_{kl}(t)$ with high probability.
\par}

More precisely, when $l$ and $t$ are given, $J_{kl}(t)$ depends only on the
variables~$X_{k'}$ with $k'\equiv k\,$ (mod~$h$) and $k'\neq l$. The unit
cube on those variables is partitioned into $g$~parts
$P_0,P_1,\ldots,P_{g-1}$ such that $J_{kl}(t)=j$ when the variables lie
in~$P_j$; the volume of~$P_j$ is $\Pr\[J_{kl}(t)=j]$. We will divide
each $P_j$ into $g$~sets $P'_{j0}$, $P'_{j1}$, \dots,~$P'_{j(g-1)}$, and
define $J_{kl}^{\ast}(t)=i$ on~$P'_{ji}$.
This subdivision, performed separately for each~$k$, will yield
independent random variables
$J_{0l}^{\ast}(t)$, $J_{1l}^{\ast}(t)$, \dots, $J_{(h-1)l}^{\ast}(t)$.
We will show that the subdivision can be done in such a way that
$$\eqalignno{\Pr\[J_{kl}^{\ast}(t)=j]&=1/g\,,&(5.1)\cr
\Pr\[J^{\ast}_{kl}(t)\neq J_{kl}(t)]&<\phi(t)\,,&(5.2)\cr}$$
for $0\leq j<g$ and $0\leq k<h$.
Thus, we will have perturbed the values of $J_{kl}(t)$
with low probability when $\phi(t)$ is small.

The following construction does what we need, and more:

\proclaim
Lemma 5. Let $p_1,\ldots,p_m$ and $p_1^{\ast},\ldots,p_m^{\ast}$ be
nonnegative real  numbers with $p_1+\cdots+p_m=p_1^{\ast}+\cdots+p_m^{\ast}=1$.
Then there are nonnegative reals $p'_{ij}$ for $1\le i,j\le m$
such that
$$\eqalignno{p_i&=\sum_{j=1}^m p'_{ij}\,,&(5.3)\cr
\noalign{\smallskip}
p_j^{\ast}&=\sum_{i=1}^m p'_{ij}\,,&(5.4)\cr}$$
and
$$
\sum_{i\neq j}p'_{ij}
=1-\sum_{j}p'_{jj}={1\over2}\sum_j|p_j-p_j^{\ast}|.
\eqno(5.5)
$$

\proof
This is a special case of ``maximal coupling'' in probability theory
[\Gold; \Lind, \S III.14]; it can be proved as follows.

Let $p'_{jj}=\min(p_j,p_j^{\ast})$, and observe that
$$
\sum_{j}p'_{jj}
=\sum_j\min(p_j,p_j^{\ast})
=\sum_j{\textstyle{1\over2}}(p_j+p_j^{\ast}-|p_j-p_j^{\ast}|)
=1-{1\over2}\,\sum_j|p_j-p_j^{\ast}|\,.
\eqno(5.6)
$$
The existence of nonnegative $p'_{ij}$, $i\neq j$, such that (5.3) and (5.4)
hold follows from the max flow--\allowbreak min cut theorem~[\FF]: Consider a
network with a source~$s$, a sink~$t$, and $2m$ nodes
$v_1,\ldots,v_m,v^{\ast}_1,\ldots,v^{\ast}_m$; the edges
are $sv_j$ with capacity $p_j-p'_{jj}$, $v^{\ast}_jt$ with
capacity $p^{\ast}_j-p'_{jj}$, and $v_iv^{\ast}_j$ with infinite
capacity.~~\pfbox

\medbreak\noindent
{\bf 6. The effect of perturbation.}\quad
When independent
random variables $J^{\ast}_{kl}(t)$ have been defined satisfying (5.1)
and (5.2), we can use them to define $Q_l^{\ast}(t)$ as in (2.3) and
$W_{jj'l}^{\ast}(t)$ as in (2.4). This value $W_{jj'l}^{\ast}(t)$ has
already been evaluated in (4.3); we want now to use the idea of perturbation
to see how much $W_{jj'l}(t)$ can differ from $W^{\ast}_{jj'l}(t)$.

Since $Q_l(t)=O(h)$ and
$$|Q_l(t)-Q^{\ast}_l(t)|\leq
\sum_{k=0}^{h-1}\,\[J_{kl}(t)\neq J^{\ast}_{kl}(t)]\,,\eqno(6.1)$$
we have
$$\eqalignno{%
\big|@W_{jj'l}(t)-W^{\ast}_{jj'l}(t)@\big|
&=\Bigl|\bigl(\Pr\[J_{ll}(t)=j]-\Pr\[J^{\ast}_{ll}(t)=j]\bigr)\,{\rm E}\,
\bigl|@Q_l(t)-\lfloor hd/g\rfloor@\bigr|\cr
\noalign{\smallskip}
&\qquad \null +\Pr\[J_{ll}^{\ast}(t)=j]\bigl({\rm E}\,\bigl|@Q_l(t)-
\lfloor hd/g\rfloor@\bigr|-{\rm E}\,\bigl|@Q_{kl}^{\ast}(t)-\lfloor
hd/g\rfloor@\bigr|\bigr)\Bigr|\cr
\noalign{\smallskip}
&<{1\over g}\,\phi(t)O(h)+{1\over g}\,\sum_{k=0}^{h-1}\,\Pr
\[J_{kl}(t)\neq J_{kl}^{\ast}(t)]\cr
\noalign{\smallskip}
&=O\left({h\over g}\right)\phi(t)\,.&(6.2)\cr}$$
(We assume that $J_{kl}^{\ast}(t)=J^{\ast}_{(k\bmod h)l}(t)$ when $k\geq
h$.)

To complete our estimate, we need to integrate this difference over all~$t$.

\proclaim
Lemma 6. $\int_0^1\phi(t)\,dt=O(g^2h/n)$.

\proof
The case $n<4h$ is trivial. Otherwise we have
$$\eqalign{\int_0^1\phi(t)\,dt
&=2\int_0^{1/2}\phi(t)\,dt\cr
\noalign{\smallskip}
&<4\int_0^{1/2}\,\sum_{k=1}^{\infty} e^{-2tk^2n/g^2h}\,dt\cr
\noalign{\smallskip}
&<4\int_0^{\infty}\,\sum_{k=1}^{\infty}e^{-2tk^2n/g^2h}\,dt\cr
\noalign{\smallskip}
&=4\,\sum_{k=1}^{\infty}\,{g^2h\over 2k^2n}={\pi^2\over 3}\,{g^2h\over n}\,.
\quad\pfbox\cr}$$

\proclaim
Theorem 1.
The average number of inversions remaining after $h$-sorting and then
$g$-sorting a random permutation of $n$~elements, when $h$ is relatively
prime to~$g$, is $\psi(h,g)@n+O(g^3h^2)$, where $\psi(h,g)$ is given by\/
$(0.2)$.

\proof
By (6.2) and Lemmas 2, 4, and~6, the average
is $\psi(h,g)@n$ plus
$$\eqalign{\sum_{\scriptstyle 0\leq j<j'<g\atop
\scriptstyle 0\leq
l<n}\,\int_0^1\bigl(W_{jj'l}(t)-W_{jj'l}^{\ast}(t)\bigr)\,dt
&=O(g^2n)O(h/g)\int_0^1\phi(t)\,dt\cr
&=O(g^3h^2)\,.\quad\pfbox\cr}$$

Notice that the proof of this theorem implicitly uses Lemma 5 for each
choice of~$l$ and~$t$, without requiring any sort of continuity between the
values of $J^{\ast}_{kl}(t)$ as $t$ varies. We could have defined
$J^{\ast}_{kl}(t)$ in a continuous fashion; indeed, the random variables
$[X_k<t]$ partition the $(n-1)$-cube into $2^{n-1}$~subrectangles in each
of which $J_{kl}(t)$ has a constant value, so we could define $J^\ast_{kl}(t)$
over $(n-1)$-dimensional rectangular prisms with
smooth transitions as a function of~$t$. But such complicated refinements
are not necessary for the validity of the perturbation argument.

\medbreak\noindent
{\bf 7. Asymptotics.}\quad
Our next goal is to estimate $\psi(h,g)$ when $h$ and $g$ are large. Notice
that
$$\psi(h,g)={1\over 2}\,\sum_{d=1}^{g-1}\,{\rm E}\,\left|
Z(h-1,d/g)-\left\lfloor{hd\over g}\right\rfloor\right|\eqno(7.1)$$
where $Z(m,p)$ has the binomial distribution with parameters $m$ and~$p$.
The mean of $Z(h-1,d/g)$ is $(h-1)d/g=\lfloor hd/g\rfloor +O(1)$, and the
variance is $(h-1)d(g-d)/g^2$. If we replace $Z$ by a normally distributed
random variable with this same mean and variance, the expected value of
$|Z-\lfloor hd/g\rfloor|$ is approximately
$(2\pi)^{-1/2}\int_{-\infty}^{\infty}|t|
e^{-t^2\!/2}\,dt=2/\sqrt{2\pi}$ times the standard deviation, so (7.1) will
be approximately
$${1\over g}\,\sqrt{h\over
2\pi}\,\sum_{d=1}^{g-1}\,\sqrt{d(g-d)}\,.\eqno(7.2)$$
The detailed calculations in the remainder of this section justify this
approximation and provide a rigorous error bound.

\proclaim
Lemma 7. If $Z$ has the binomial distribution with parameters $(m,p)$,
and $\lfloor mp\rfloor\le a\le \lceil mp\rceil$, then
$${\rm E}\,|Z-a|=\sqrt{{2p(1-p)m\over\pi}}
+O\biggl({1\over\sqrt{mp(1-p)}}\biggr)\,.\eqno(7.3)$$

\proof
Consider first the case $a=mp$.
By a formula of De Moivre~[\DeM, page~101] and Poincar\'e~[\Poin, pages
56--60],  see Diaconis and Zabell~[\DZ],
$${\rm E}\,|Z-mp@|=2\lceil mp\rceil{m\choose\lceil mp\rceil}
p^{\lceil mp\rceil}(1-p)^{m+1-\lceil mp\rceil}.
\eqno(7.4)$$
In order to prove (7.3) in this case we may assume that $p\le1/2$, since
$|Z-mp@|=|m-Z-m(1-p)|$. Moreover, we may assume that
$mp>1$ since (7.3) otherwise is trivial.
Then, a routine application of
Stirling's approximation shows that
$${\rm E}\,|Z-mp@|=\sqrt{{2p(1-p)m\over\pi}}
\exp{\biggl(O\Bigl({1\over mp}\Bigr)\biggr)}\,.\eqno(7.5)$$
Next observe that if $\lfloor mp\rfloor\le a\le \lceil mp\rceil$, we have
$${\rm E}\,|Z-a|=
{\rm E}\,|Z-mp@|+(mp-a)\bigl(1-2\Pr\[Z\le mp]\bigr)\,.
\eqno(7.6)$$
Since $\Pr\[Z\le mp]={1\over2}+O\bigl((mp(1-p))^{-1/2}\bigr)$,
for example by the Berry--Esseen estimate
of the error in the central limit
theorem~[\Fii, \S XVI.5], the result follows.~~\pfbox

\proclaim
Corollary. The asymptotic value of $\psi(h,g)$ is
$$\psi(h,g)=\sqrt{{\pi h\over 128}}\,g+O(g^{-1/2}h^{1/2})+O(gh^{-1/2})
\,.\eqno(7.7)$$

\proof
Since $\lfloor hd/g\rfloor\le\lfloor(h+1)d/g\rfloor\le\lfloor(hd+g-1)/g\rfloor
=\lceil hd/g\rceil$,
Lemma 7 yields
$$\eqalign{\psi(h+1,g)
&={1\over2}\sum_{d=1}^{g-1}\,{\rm E}\,\left|Z(h,d/g)-
  \left\lfloor{(h+1)d\over g}\right\rfloor\right|\cr
&=\sum_{d=1}^{g-1}\left(\sqrt{{h\over2\pi}{d\over g}\Bigl(1-{d\over g}\Bigr)}
+O\left(\left(h{d\over g}\Bigl(1-{d\over g}\Bigr)\right)^{\!-1/2}\right)\right)
\cr
&=\sqrt{h\over2\pi}\,
 \sum_{d=1}^{g-1}\sqrt{{d\over g}\Bigl(1-{d\over g}\Bigr)}
+O(gh^{-1/2})\,.\cr}$$
And Euler's summation formula with $f(x)=\sqrt{(x/g)(1-x/g)}$ tells us that
$$\eqalign{\sum_{d=1}^{g-1}f(d)
&=\int_1^{g-1}f(x)\,dx+{1\over2}f(1)+{1\over2}f(g-1)+{1\over12}f'(g-1)
 -{1\over12}f'(1)-R\cr
&=g\int_0^1\sqrt{t(1-t)}\,dt+O(g^{-1/2})={\pi g\over8}+O(g^{-1/2})\cr}$$
because
$$|R|=\left|\int_1^{g-1}{B_2(x\bmod1)\over2}f''(x)\,dx\right|
\le{1\over12}\int_1^{g-1}\bigl|f''(x)\bigr|\,dx=
{1\over12}f'(1)-{1\over12}f'(g-1)\,.\quad
\pfbox$$

The error term is thus $O(g^{1/2})$
when $h=g^2+1$; for example, we have
$$\vcenter{\halign{\hfil#\qquad&\hfil#\qquad&\hfil#\qquad&\hfil#\qquad%
&\hfil#\cr
$h$\hfil&$g$\hfil&$\psi(h,g)$\hfil&$\sqrt{\pi h/128}\,g$&
 difference/$\sqrt g$\cr
\noalign{\smallskip}
901&30&140.018&141.076&0.1933\cr
1601&40&249.539&250.741&0.1900\cr
2501&50&390.412&391.739&0.1877\cr
}}$$

\medbreak\noindent
{\bf 8. Common factors.}\quad
Now let's consider the behavior of shellsort with increments $(ch,cg,1)$,
where $c$ is an integer $>1$. It is easy to see that the first two passes
are equivalent to the first two passes of $(h,g,1)$ shellsort on $c$
independent subarrays $(X_a,X_{a+c},X_{a+2c},\ldots)$ of size
$\lceil(n-a)/c\rceil$ for $0\leq a<c$. The inversions that remain are the
$\psi(h,g)@n+O(g^3h^2c)$ inversions within these subarrays, plus
``cross-inversions'' between ${c\choose 2}$ pairs of subarrays.

Yao
[\Yao, Theorem 2]
proved that the average number of cross-inversions is ${1\over
8}\,\sqrt{\pi c}\,(1-c^{-1})@n^{3/2}+O(cghn)$. The following lemma improves
his error term slightly.

\proclaim
Lemma 8. The average
number of cross-inversions after $ch$-sorting and $cg$-sorting is
$${1\over 8}\,\sqrt{\pi c}\,(1-c^{-1})@n^{3/2}+O(cgh^{1/2}n)+O(c^2g^3h^2)\,.
\eqno(8.1)$$

\proof
Let's consider first the process of $h$-sorting and $g$-sorting two
independent arrays $(X_0,X_1$,
$\ldots,X_{n-1})$ and
$(\widehat{X}_0,\widehat{X}_1,\ldots,\widehat{X}_{n-1})$, then interleaving
the results to obtain $(X''_0,\widehat{X}''_0,X''_1,\widehat{X}''_1,
\ldots\,$
$X''_{n-1},\widehat{X}''_{n-1})$. The cross inversions are then the
pairs $\{X''_l,\widehat{X}''_{l'}\}$ where either
$X''_l>\widehat{X}''_{l'}$ and $l\leq l'$ or $X''_l<\widehat{X}''_{l'}$ and
$l>l'$.

Recasting this process in the model of section 2 above, we assume that
$X_l=t$, while the other $2n-1$ variables
$(X_0,\ldots,X_{l-1},\ldots,X_{n-1},\widehat{X}_0,\ldots,\widehat{X}_{n-1})$
are independent and uniformly distributed between 0 and~1. We define
$$Y_{kl}(t)=\sum_{\scriptstyle k'\equiv k\,(\bmod\;h)\atop
\scriptstyle 0\leq k'<n}\,\[X_{k'}<t]\,,\quad
\widehat{Y}_{kl}(t)=\sum_{\scriptstyle k'\equiv k\,(\bmod\;h)\atop
\scriptstyle 0\leq k'<n}\,\[\widehat{X}_{k'}<t]\eqno(8.2)$$
as in (2.1). The elements of each array are divided into $h$ subarrays by
$h$-sorting, and the elements $<t$ have $Y_{kl}(t)$ and
$\widehat{Y}_{kl}(t)$ elements in the $k$\/th subarrays. Then $g$-sorting
will form $g$~lists, with
$$L_{jl}(t)=\sum_{k=0}^{h-1}\,\left\lceil{Y_{kl}(t)-a_{kj}\over g}
\right\rceil\eqno(8.3)$$
elements $<t$ in the $j$\/th list of the first array,
where $a_{kj}\in\{0,1,\ldots,g-1\}$ is given by
$k+a_{kj}h \equiv j$ (mod~$g$).
Similarly, there will be
$$\widehat{L}_{jl}(t)=\sum_{k=0}^{h-1}\,
\left\lceil{\widehat{Y}_{kl}(t)-a_{kj}\over g}
\right\rceil\eqno(8.4)$$
elements $<t$ in the $j$\/th list of the second. Element $X_l=t$ of the
first array will go into list $j=J_{ll}(t)$ as before, where $J_{kl}(t)$ is
defined in (2.2). The number of cross-inversions between this element and
the elements of the second array will then be
$$V_l(t)=\sum_{j'=0}^{g-1}\,\bigl|@\widehat{L}_{j'l}(t)-L_{jl}(t)-\[j'<j]
@\bigr|\,.\eqno(8.5)$$
The average total number of cross-inversions is the sum of ${\rm E}\,V_l(t)$
over all~$l$, integrated for $0\leq t\leq 1$.

We know from Lemma 3 that the numbers $Y_{kl}(t)\bmod g$
have approximately a uniform distribution. Therefore
$$\left\lceil{Y_{kl}(t)-a_{kj}\over g}\right\rceil =
{Y_{kl}(t)-a_{kj}+R_{jkl}(t)\over g}$$
where $R_{jkl}(t)$ is approximately uniform on $\{0,1,\ldots,g-1\}$. It
follows that
$$L_{jl}(t)={Z_l(t)\over g}+\sum_{k=0}^{h-1}\,
\left({R_{jkl}(t)-a_{kj}\over g}\right)\,,\eqno(8.6)$$
where
$$Z_l(t)=\sum_{k=0}^{h-1}\,Y_{kl}(t)$$
is the total number of elements in the first array that are $<t$.

Since $R_{jkl}(t)$ depends on $Y_{kl}(t)\bmod g$ only, or equivalently on
$J_{kl}(t)$,
we may use the perturbed truly uniform random variables
$J^{\ast}_{kl}(t)$ in section 5
(or repeat the argument there with $R_{jkl}(t)$)
and construct random variables $R_{jkl}^{\ast}(t)$ that are
uniform on $\{0,1,\ldots,g-1\}$ and satisfy
$\Pr\[R_{jkl}^{\ast}(t)\neq R_{jkl}(t)]<\phi(t)$;
moreover, the variables $R_{jkl}^{\ast}(t)$ are independent for $0\le k<h$
and fixed $j$ and $l$.
Consequently
$${\rm E}\,|R_{jkl}^{\ast}(t)-R_{jkl}(t)|
\le g\Pr\[R_{jkl}^{\ast}(t)\neq R_{jkl}(t)]<g\phi(t)\,.
\eqno(8.7)$$
By independence and the fact that ${\rm E}\,R_{jkl}^{\ast}(t)=(g-1)/2$,
$${\rm E}\,\left(@\sum_{k=0}^{h-1}R_{jkl}^{\ast}(t)-h(g-1)/2\right)^2
=\sum_{k=0}^{h-1}{\rm E}\,\bigl(R_{jkl}^{\ast}(t)-(g-1)/2\bigr)^2
<hg^2,$$
which by the Cauchy--Schwarz inequality yields
$${\rm E}\,\left|@\sum_{k=0}^{h-1}R_{jkl}^{\ast}(t)-h(g-1)/2@\right|
<\sqrt{h}g.
\eqno(8.8)$$

Let $W_{jl}={1\over g}\bigl(\sum_{k=0}^{h-1}R_{jkl}(t)-h(g-1)/2\bigr)$
and $b_j={1\over g}\bigl(h(g-1)/2-\sum_{k=0}^{h-1}a_{kj}\bigr)$; then
$$L_{jl}(t)={Z_l(t)\over g}+W_{jl}+b_j,\eqno(8.9)$$
where by (8.7) and (8.8)
$${\rm E}\,|W_{jl}(t)|<\sqrt{h}+h\phi(t).$$

A similar argument shows that
$$\widehat{L}_{jl}(t)={\widehat{Z}_l(t)\over g}+
\widehat{W}_{jl}+b_j.$$
Hence
$$V_l(t)
=\sum_{j'=0}^{g-1}\,\left({|\widehat{Z}_l(t)-Z_l(t)|\over g}
+O(|W_{j'l}|+|\widehat{W}_{j'l}|+1)\right)$$
and
$${\rm E}\,V_l(t)
={\rm E}\,|\widehat{Z}_l(t)-Z_l(t)|
+O(g\sqrt{h}\,)+O(gh)@\phi(t)\,.\eqno(8.10)$$
The quantity $|\widehat{Z}_l(t)-Z_l(t)|$ is just what we would get
if we were counting the cross-inversions between two fully sorted arrays
that have been interleaved. Therefore
$$\int_0^1\sum_{l=0}^{n-1}\,{\rm E}\,\bigl|\widehat{Z}_l(t)-Z_l(t)
\bigr|\,dt$$
must be the average number of inversions of a random 2-ordered permutation
of $2n$~elements; this, according to Douglas~H. Hunt in 1967, is exactly
$n@2^{2n-2}\!\left/{2n\choose n}\right.$
[\Kiii, exercise 5.2.1--14].
Since ${2n\choose n}= \bigl(1+O(1/n)\bigr)4^n/\sqrt{\pi n}$,
we obtain the desired total
$$\int_0^1\,{\rm E}\,\sum_{l=0}^{n-1}\,V_l(t)\,dt = {\sqrt{\pi}\,n^{3/2}\over 4}
+O(gh^{1/2}n)+O(g^3h^2)\eqno(8.11)$$
by Lemma 6.
Similarly, the same result holds for two arrays of different sizes $n+O(1)$.

Lemma 8 follows if we replace $n$ by $n/c+O(1)$ in (8.11) and multiply by
${c\choose 2}$.~~\pfbox

\medbreak\noindent
{\bf 9. The total cost.}\quad
So far we have been considering only the number of inversions removed
during the third pass of a three-pass shellsort. But the first two passes
can be analyzed as in Yao's paper~[\Yao]:

\proclaim
Theorem 2. Let $g$ and $h$ be relatively prime and let $c$ be a positive
integer. The average number of inversions removed when
$(ch,cg,1)$-shellsort is applied to a random $n$-element array~is
$${n^2\over 4ch}+O(n)\eqno(9.1)$$
on the first pass,
$${1\over 8g}\,\sqrt{{\pi\over ch}}\,(h-1)@n^{3/2}+O(hn)\eqno(9.2)$$
on the second, and
$$\psi(h,g)@n+{1\over 8}\,\sqrt{{\pi\over
c}}\,(c-1)@n^{3/2}+O\bigl((c-1)gh^{1/2}n\bigr)+O(c^2g^3h^2)\eqno(9.3)$$
on the third.

\proof
The first pass removes an average of ${1\over 4}\bigl(n/ch+O(1)\bigr)^2$
inversions from~$ch$ subarrays of size $\lfloor n/ch\rfloor$ or $\lceil
n/ch\rceil$; this proves (9.1). The second pass is equivalent to the second
pass of $(h,g,1)$-shellsort on $c$~independent subarrays of sizes $\lfloor
n/c\rfloor$  or $\lceil n/c\rceil$. Equation (9.3) is Lemma~8. So the
theorem will follow if we can prove (9.2) in the case $c=1$.
And that case follows from
[\Yao, equation (32)],
with the $O(n)$ term replaced by $O(n/kh)$ in the notation of that paper.
(See also
[\Kiii, second edition, exercise 5.2.1--40.)~~\pfbox

\proclaim
Corollary. If $h=\Theta(n^{7/15})$, $g=\Theta(n^{1/5})$, and $\gcd(g,h)=1$,
the running time of $(h,g,1)$-shellsort is $O(n^{23/15})$.

\proof
The first pass takes time $O(n^{2-7/15})$, by (9.1); the second takes
$O(n^{3/2+7/30-1/5})+O(n^{1+7/15})$, by (9.2); and the third takes
$O(n^{1+1/5+7/30})+O(n^{3/5+14/15})$ by (7.6) and (9.3).~~\pfbox

\medbreak\noindent
{\bf 10. Two conjectures.}\quad
Our estimate $O(g^3h^2)$ for the difference between $\psi(h,g)@n$ and the
average number of third-pass inversions may not be the best possible. In
fact, the authors conjecture that the difference is at most
$O(g^3h^{3/2})$. This sharper bound may perhaps follow from methods
analogous to those in the proof of Lemma~8.

If such a conjecture is valid, the running time of $(h,g,1)$-shellsort will
be $O(n^{3/2})$ when $h\approx n^{1/2}$ and $g\approx n^{1/4}$. A~computer
program was written to test this hypothesis by applying $(h,g,1)$-shellsort
to random arrays of $n$~elements with $h=g^2+1$ and $n=g^2h=g^4+g^2$. The
following empirical results were obtained, to three significant figures:
$$\vcenter{\halign{%
\hfil#\quad&\hfil$#$\hfil\quad&\hfil#\qquad\qquad
&\hfil#\quad&\hfil$#$\hfil\quad&\hfil#\cr
$g$&\hbox{inversions}&$\psi(h,g)@n$%
&$g$&\hbox{inversions}/10^5&$\psi(h,g)@n/10^5$\cr
\noalign{\smallskip}
1&0\pm 0&0&17&36.6\pm 2.36/32&37.3\cr
2&7.12\pm 2.09/100&7.5&18&51.7\pm 3.35/32&52.6\cr
3&94.4\pm 13.6/100&98.3&19&71.5\pm 4.81/32&72.9\cr
4&563\pm 59.1/100&581&20&97.3\pm 6.14/10&99.2\cr
5&2210\pm 195/100&2280&21&130\pm 8.93/10&133\cr
6&6740\pm 560/100&6910&22&174\pm 12.3/10&176\cr
7&17200\pm 1300/100&17600&23&226\pm 14.0/10&230\cr
8&38600\pm 2820/100&39500&24&291\pm 16.8/10&297\cr
9&78900\pm 5670/100&80600&25&368\pm 23.7/10&380\cr
10&149000\pm 10600/100&152000&26&475\pm 29.1/10&480\cr
11&265000\pm 17200/32&271000&27&595\pm 39.0/10&603\cr
12&447000\pm 30300/32&458000&28&735\pm44.9/10&750\cr
13&727000\pm 49300/32&742000&29&922\pm52.1/10&926\cr
14&1140000\pm 75400/32&1160000&30&1110\pm 74.0/10&1140\cr
15&1730000\pm 116000/32&1760000&31&1370\pm 97.9/10&1380\cr
16&2530000\pm 166000/32&2590000&32&1650\pm 101/10&1670\cr}}$$
(The inversion counts are given here in the form $\mu\pm\sigma/\sqrt{r}$,
where $\mu$ and~$\sigma$ are the empirical mean and standard derivation in
$r$~independent trials. For example, 10000 trials were made when $g\leq
10$, but only 100 trials were made when $g\geq 20$.) Both mean and standard
derivation seem to be growing proportionately to $g^6\approx n^{3/2}$, with
$\sigma\approx\mu/15$ for $g\geq 10$.

These data suggest also another conjecture, that the average number of
inversions is $\leq\psi(h,g)@n$ when $h$ and~$g$ are relatively prime.
Indeed, the deviations from uniformity between ${\cal A}$ and~${\cal
A}^{\ast}$ should tend to cause fewer inversions, because ${\cal A}$ forces
the balance condition $Y_{kl}(1) = n/h+O(1)$ for all $k$ and~$l$. This
second conjecture obviously implies running time $\Theta(n^{3/2})$ when
$h=\Theta(n^{1/2})$ and $g=\Theta(n^{1/4})$.

\medbreak\noindent
{\bf 11. More than three increments?}\quad
It may be possible to extend this analysis to $(h,g,f,1)$-shellsort, by
analyzing the following stochastic algorithm. ``Initialize two sets of
counters $(I_0,I_1,\allowbreak
\ldots,I_{g-1})$ and $(J_0,J_1,\ldots,J_{h-1})$ by setting
$I_j\leftarrow j\bmod f$ and $J_k=k\bmod g$ for all $j$ and~$k$. Then
execute the following procedure $n$~times: Choose a random~$k$ in the range
$0\leq k<h$. Set $j\leftarrow J_k$ and $i\leftarrow I_j$; then set
$J_k\leftarrow (J_k+h)\bmod g$ and $I_j\leftarrow (I_j+g)\bmod f$.''

Consider the transition from $l=l_t$ to $l'=l_{t+1}$ in the proof of
Lemma~1. When elements enter the array in increasing order, the choice
of~$k$ represents the subarray that will contain a new element~$X$ during
the $h$-sort; then $X$ goes into list~$j$ during the $g$-sort, and into
list~$i$ during the $f$-sort.
We can therefore obtain the contribution of~$X$ to the inversions between
lists~$i$ and~$i'$ for $i<i'<f$, by considering a state~$P_l$ obtained from
the $I$~table just as $Q_l$ was obtained from the $J$~table in Lemma~1.

\bigskip
\centerline{\bf References}

\bib
[\DeM]
Abraham De Moivre, {\sl Miscellanea Analytica de Seriebus et Quadraturis},
(London: J. Tonson and J. Watts, 1730).

\bib
[\DZ]
Persi Diaconis and Sandy Zabell, ``Closed form summation for classical
distributions: Variations on a theme of De Moivre,''
{\sl Statistical Science\/ \bf 6} (1991), 284--302.

\bib
[\Fii]
William Feller, {\sl An Introduction to Probability Theory and Its
Applications\/ \bf2} (New York: Wiley, 1966).

\bib
[\FF]
L. R. Ford, Jr., and D. R. Fulkerson, ``Maximal flow through a network,'' {\sl
Canadian Journal of Mathematics\/ \bf8} (1956), 399--404.

\bib
[\Gold]
Sheldon Goldstein, ``Maximal coupling,'' {\sl Zeitschrift f\"ur
Wahrscheinlichkeitstheorie und verwandte Gebiete\/ \bf46} (1979), 193--204.

\bib
[\Kii]
Donald E. Knuth, {\sl Seminumerical Algorithms}, Volume~2 of {\sl The
Art of Computer Programming\/} (Reading, Massachusetts: \AW, 1969).
Second edition, 1981.

\bib
[\Kiii]
Donald E. Knuth, {\sl Sorting and Searching}, Volume~3 of {\sl The
Art of Computer Programming\/} (Reading, Massachusetts: \AW, 1973).
Second edition, 1997.

\bib
[\KMP]
Donald E. Knuth, Rajeev Motwani, and Boris Pittel, ``Stable husbands,''
{\sl Random Structures and Algorithms\/ \bf 1} (1990), 1--14.

\bib
[\Lind]
Torgny Lindvall, {\sl Lectures on the Coupling Method\/} (New York: Wiley,
1992).

\bib
[\Poin]
Henri Poincar\'e, {\sl Calcul des Probabilit\'es\/} (Paris: Georges Carr\'e,
1896).

\bib
[\Yao]
Andrew Chi-Chih Yao, ``An analysis of $(h,k,1)$-Shellsort,''
{\sl Journal of Algorithms\/ \bf 1} (1980), 14--50.

\bigskip\bigskip
\noindent
Authors' addresses:

\nobreak\noindent
Svante Janson, Department of Mathematics, Uppsala University,
P.O.Box 480, 75106 Uppsala, Sweden; {\tt
svante.janson{\char'100}math.uu.se}

\noindent
Donald E. Knuth, Computer Science Department, Gates Building 4B, Stanford
University, Stanford CA 94305--9045 USA;
{\tt http://www-cs-faculty.stanford.edu/{\char'176}knuth}

\bye